\documentclass[10pt,conference]{IEEEtran}
\usepackage{cite}
\usepackage{amsmath,amssymb,amsfonts}
\usepackage{algorithmic}
\usepackage{graphicx}
\usepackage{textcomp}
\usepackage{xcolor, colortbl}
\usepackage{xspace}
\usepackage{multirow}
\usepackage{caption}
\usepackage{subcaption}
\usepackage{footnote}
\usepackage{url}
\usepackage{listings}
\usepackage{tcolorbox}
\usepackage{booktabs}
\usepackage{enumitem}
\usepackage{fixltx2e}
\usepackage{multicol}
\usepackage{balance}
\usepackage{listings}
\usepackage[normalem]{ulem}

\newcommand{\ie}{\emph{i.e.,}\xspace}
\newcommand{\eg}{\emph{e.g.,}\xspace}
\newcommand{\etc}{etc.\xspace}
\newcommand{\etal}{\emph{et~al.}\xspace}

\newcommand{\figref}[1]{Fig.~\ref{#1}\xspace}

\newcommand{\tabref}[1]{Table~\ref{#1}\xspace}

\newcommand{\RQ}[1]{RQ$_{\textbf{#1}}$\xspace}

\newlength\Linewidth
\def\findlength{
	\setlength\Linewidth\linewidth
	\addtolength\Linewidth{-4\fboxrule}
	\addtolength\Linewidth{-3\fboxsep}
}

\newenvironment{summarybox}
	{\par\begingroup
	\setlength{\fboxsep}{5pt}\findlength
	\setbox0=\vbox\bgroup\noindent
	\hsize=0.95\linewidth
	\begin{minipage}{0.95\linewidth}\normalsize}
	{\end{minipage}\egroup
	   \vspace{3pt}
	\textcolor{gray}{\fboxsep1.5pt\fbox{\fboxsep5pt\colorbox{white}{\normalcolor\box0}}}
	\endgroup\par\noindent
	\normalcolor\ignorespacesafterend}

\makeatletter
\newcommand\footnoteref[1]{\protected@xdef\@thefnmark{\ref{#1}}\@footnotemark}
\makeatother

\newboolean{showcomments}
\setboolean{showcomments}{true}
\ifthenelse{\boolean{showcomments}}
  {\newcommand{\nb}[2]{
    \fbox{\bfseries\sffamily\scriptsize#1}
    {\sf\small$\blacktriangleright$\textit{#2}$\blacktriangleleft$}
   }
  }
  {\newcommand{\nb}[2]{}
  }

\newcommand\codeword[1]{\texttt{\textcolor{black}{#1}}}
\newcommand\placeholder[1]{$\left\{ \textrm{\textit{{#1}}} \right\}$}

\def\BibTeX{{\rm B\kern-.05em{\sc i\kern-.025em b}\kern-.08em
    T\kern-.1667em\lower.7ex\hbox{E}\kern-.125emX}}
\begin{document}

\title{Fixing Dockerfile Smells: An Empirical Study$^{\scalebox{.8}{$\scriptscriptstyle *$}}$}

\author{\IEEEauthorblockN{Giovanni Rosa}
\IEEEauthorblockA{\textit{STAKE Lab} \\
\textit{University of Molise}\\
Pesche, Italy \\
\emph{giovanni.rosa@unimol.it}}
\and
\IEEEauthorblockN{Simone Scalabrino}
\IEEEauthorblockA{\textit{STAKE Lab} \\
\textit{University of Molise}\\
Pesche, Italy \\
\emph{simone.scalabrino@unimol.it}}
\and
\IEEEauthorblockN{Rocco Oliveto}
\IEEEauthorblockA{\textit{STAKE Lab} \\
\textit{University of Molise}\\
Pesche, Italy \\
\emph{rocco.oliveto@unimol.it}}
}

\maketitle

\begin{abstract}
\textit{Background.} 
Containerization technologies are widely adopted in the DevOps workflow. The most commonly used one is Docker, which requires developers to define a specification file (Dockerfile) to build the image used for creating containers. There are several best practice rules for writing Dockerfiles, but the developers do not always follow them. Violations of such practices, known as Dockerfile smells, can negatively impact the reliability and the performance of Docker images. Previous studies showed that Dockerfile smells are widely diffused, and there is a lack of automatic tools that support developers in fixing them. However, it is still unclear what Dockerfile smells get fixed by developers and to what extent developers would be willing to fix smells in the first place.

\textit{Objective.} The aim of our exploratory study is twofold. First, we want to understand what Dockerfiles smells receive more attention from developers, \ie are fixed more frequently in the history of open-source projects. Second, we want to check if developers are willing to accept changes aimed at fixing Dockerfile smells (\eg generated by an automated tool), to understand if they care about them.

\textit{Method.} In the first part of the study, we will evaluate the survivability of Dockerfile smells on a state-of-the-art dataset composed of 9.4M unique Dockerfiles. We rely on a state-of-the-art tool (\textit{hadolint}) for detecting which Dockerfile smells disappear during the evolution of Dockerfiles, and we will manually analyze a large sample of such cases to understand if developers fixed them and if they were aware of the smell.
In the second part, we will detect smelly Dockerfiles on a set of GitHub projects, and we will use a rule-based tool to automatically fix them. Finally, we will open pull requests proposing the modifications to developers, and we will quantitatively and qualitatively evaluate their outcome.
\end{abstract}


\noindent{\footnotesize *Note: This study was accepted at the ICSME 2022 Registered Reports Track.}


\section{Introduction}
Software systems are developed to be deployed and used. Operating software in a production environment, however, entails several challenges. Among the others, it is very important to make sure that the software system behaves exactly as in a development environment. Virtualization and, above all, containerization technologies are increasingly being used to ensure that such a requirement is met\footnote{\url{https://portworx.com/blog/2017-container-adoption-survey/}}. To this end, Docker\footnote{\url{https://www.docker.com/}} is one of the most popular platforms used in the DevOps workflow: It is the main containerization framework in the open-source community \cite{cito2017empirical}, and is widely used by professional developers\footnote{\label{stacksurvey}\url{https://insights.stackoverflow.com/survey/2021}}. 
Also, Docker is the most loved and most wanted platform in the 2021 StackOverflow survey\footnoteref{stacksurvey}. Docker allows releasing applications together with their dependencies through containers (\ie virtual environments) sharing the host operating system kernel. 
Each Docker image is defined through a Dockerfile, which contains instructions to build the image containing the application.
All the Docker images are hosted on an online repository called DockerHub \footnote{\url{https://hub.docker.com/}}. 
Since its introduction in 2013, Docker counts 3.3M of Desktop installations, and 318B image pulls from DockerHub\footnote{\url{https://www.docker.com/company/}}. 
Defining Dockerfiles, however, is far from trivial: Each application has its own dependencies and requires specific configurations for the execution environment. Previous work \cite{wu2020characterizing} introduced the concept of Dockerfile smells, which are violations of best practices, similarly to code smells \cite{becker1999refactoring}, and a catalogue of such problems\footnote{\url{https://github.com/hadolint/hadolint/wiki}}. 
The presence of such smells might increase the risk of build failures, generate oversized images, and security issues \cite{cito2017empirical, zhang2018one,henkel2020learning,zerouali2019relation}. Previous work studied the prevalence of Dockerfile smells \cite{cito2017empirical, lin2020large, eng2021revisiting}. 
Despite the popularity and adoption of Docker, there is still a lack of tools to support developers in improving the quality and reliability of containerized applications, as tools for automatic refactoring of code smells on Dockerfiles \cite{ksontini2021refactorings}. Relevant studies in this area investigated the prevalence of Dockerfile smells in open-source projects \cite{cito2017empirical, wu2020characterizing, lin2020large, eng2021revisiting}, the diffusion technical debt \cite{azuma2022empirical}, and the refactoring operations typically performed by developers \cite{ksontini2021refactorings}. 
While it is clear which Dockerfile smells are more frequent than others, it is still unclear which smells are more important to developers. A previous study by Eng \etal \cite{eng2021revisiting} reported how the number of smells evolve in time. Still, there is no clear evidence showing that (i) developers fix Dockerfile smells (\eg they do not disappear incidentally), and that (ii) developers would be willing to fix Dockerfile smells in the first place.

In this paper, we propose a study to fill this gap. First, we want to analyze the survivability of Dockerfile smells, to check how developers fix them, so that we can understand which ones are more relevant to them. This, however, only tells a part of the story: Developers might not correct some smells because they are harder to fix. Therefore, we also evaluate to what extent developers are willing to accept fixes to smells when they are proposed to them (\eg by an automated tool). The context of our study is represented by a state-of-the-art dataset containing about 9.4M unique Dockerfiles, along with their change history. 
For each instance of such a dataset (which is a Dockerfile snapshot), we have the list of Dockerfile smells detected with the \textit{hadolint} tool \cite{webhadolint}. The tool performs a rule check on a parsed AST representation of the input Dockerfile, based on the Docker \cite{webbestpractice} and shell script \cite{webshellcheck} best practices. 
For each Dockerfile, we will manually check a sample of the commits that make one or more smells disappear. We will aim at understanding (i) if the fix was real (\eg the smell was not removed incidentally), and (ii) if it was \textit{informed} (\eg if developers explicitly mention such an operation in the commit message). 
Then, we will then evaluate to what extent developers are willing to accept changes aimed at fixing smells. To this aim, we defined a rule-based prototype tool that automatically fixes 8 of the most frequent Dockerfile smells. We will run it on Dockefiles containing smells that it can fix and submit pull requests to developers of selected repositories. In the end, we will check how many of them get accepted for each smell type and the developers' reactions.
To summarize, the contributions that we will provide with our study are the following:

\begin{enumerate}
    \item A detailed analysis on the survivability of Dockerfile smells;

    \item An evaluation via pull requests of the willingness of developers of accepting changes aimed at fixing Dockerfile smells.
\end{enumerate}


\section{Background and Related Work}
\label{sec:related}

Technical debts \cite{cunningham1992wycash} have a negative impact on the software maintainability. A symptom of technical debts is represented by \textit{code smells} \cite{becker1999refactoring}. Code smells are poor implementation choices, that does not follow design and coding best practices, such as design patterns. They can negatively impact the maintainability of the overall software system. Mainly, code smells are defined for object-oriented systems. Some examples are duplicated code or god class (\ie a class having too much responsibilities). In the following, we first introduce smells that affect Dockerfile, and then we report recent studies on their diffusion and the practices used to improve Dockerfile quality.

\textbf{Dockerfile smells.} 
Docker reports an official list of best practices for writing Dockerfiles \cite{webbestpractice}. Such best practices also include indications for writing shell script code included in the \codeword{RUN} instructions of Dockerfiles. For example, the usage of the instruction \codeword{WORKDIR} instead of the bash command \codeword{cd} to change directory. This because each Docker instruction defines a new layer at the time of build. The violation of such practices lead to the introduction of Dockerfile smells. In fact, with Dockerfile smells, we indicate that instructions of a Dockerfile that violate the writing best practices and thus can negatively affect the quality of them \cite{wu2020characterizing}. The presence of Dockerfile smells can also have a direct impact on the behavior of the software in a production environment. For example, previous work showed that missing adherence to best practices can lead to security issues \cite{zerouali2019relation}, negatively impact the image size \cite{henkel2020learning}, increase build time and affect the reproducibility of the final image (\ie build failures) \cite{cito2017empirical,zhang2018one,henkel2020learning}. For example, the version pinning smell, that consists in missing version number for software dependencies, can lead to build failures as with dependencies updates the execution environment can change. There are several tools that support developers in writing Dockerfiles. An example is the \textit{binnacle} tool, proposed by Henkel \etal \cite{henkel2020learning} that performs best practices rule checking defined on the basis of a dataset of Dockerfiles written by experts. The reference tool used in literature for the detection of Dockerfile smells is \textit{hadolint} \cite{webhadolint}. Such a tool checks a set of best practices violations on a parsed AST version of the target Dockerfile using a rule-based approach. Hadolint detects two main categories of issues: Docker-related and shell-script-related. The former affect Dockerfile-specific instructions (\eg the usage of absolute path in the \codeword{WORKDIR} command\footnote{\url{https://github.com/hadolint/hadolint/wiki/DL3000}}). They are identified by a name having the prefix \textit{DL} followed by a number. The shell-script-related violations, instead, specifically regard the shell code in the Dockerfile (\eg in the \codeword{RUN} instructions). Such violations are a subset of the ones detected by the \textit{ShellCheck} tool \cite{webshellcheck} and they are identified by the prefix \textit{SC} followed by a number. It is worth saying that these rules can be updated and changed during time. For example, as the instruction \codeword{MAINTAINER} has been deprecated, the rule DL4000 that previously check for the usage of that instructions that was a best practice, has been updated as the avoidance of that instruction because it is deprecated.

\textbf{Diffusion of Dockerfile smells.} A general overview of the diffusion of Dockerfile smells was proposed by Wu \etal \cite{wu2020characterizing}. They performed and empirical study on a large dataset of 6,334 projects to evaluate which Dockerfile smells occurred more frequently, along with coverage, distribution and a particular focus on the relation with the characteristics of the project repository. They found that nearly 84\% of GitHub projects containing Dockerfiles are affected by Dockerfile smells, where the Docker-related smells are more frequent that the shell-script smells.
Also in this direction, Cito \etal \cite{cito2017empirical} performed an empirical study to characterize the Docker ecosystem in terms of quality issues and evolution of Dockerfiles. They found that the most frequent smell regards the lack of version pinning for dependencies, that can lead to build fails.
Lin \etal \cite{lin2020large} conducted an empirical analysis of Docker images from DockerHub and the git repositories containing their source code. They investigated different characteristics such as base images, popular languages, image tagging practices and evolutionary trends. The most interesting results are those related to Dockerfile smells prevalence over time, where the version pinning smell is still the most frequent. On the other hand, smells identified as DL3020 (\ie \codeword{COPY/ADD} usage), DL3009 (\ie clean apt cache) and DL3006 (\ie image version pinning) are no longer as prevalent as before. Furthermore, violations DL4006 (\ie usage of \codeword{RUN} pipefail) and DL3003 (\ie usage of \codeword{WORKDIR}) became more prevalent.
Eng \etal \cite{eng2021revisiting} conducted an empirical study on the largest dataset of Dockerfiles, spanning from 2013 to 2020 and having over 9.4 million unique instances. They performed an historical analysis on the evolution of Dockerfiles, reproducing the results of previous studies on their dataset. Also in this case, the authors found that smells related to version pinning (\ie DL3006, DL3008, DL3013 and DL3016) are the most prevalent. In terms of Dockerfile smell evolution, they show that the count of code smells is slightly decreasing over time, thus hinting at the fact that developers might be interested in fixing them. Still, it is unclear the reason behind their disappearance, \eg if developers actually fix them or if they get removed incidentally.



\newcommand{\rqOne}{\textit{How do developers fix Dockerfile smells?}\xspace}
\newcommand{\rqTwo}{\textit{Which Dockerfile smells are developers willing to address?}\xspace}

\section{Research Questions}

The \textit{goal} of the study that we propose is to understand whether developers are interested in fixing Dockerfile smells. The \textit{perspective} is of researchers interested in the improvement of Dockerfile quality. The \textit{context} consists of about 9.4 million of Dockerfiles, from the the largest and most recent dataset of Dockerfiles from the literature \cite{eng2021revisiting}. 

Our study is steered by the following research questions: 

\begin{itemize}
    \item \textbf{\RQ{1}:} \rqOne 
    We want to conduct a comprehensive analysis on the survivability of Dockerfile smells. Thus, we investigate what smells are fixed by developers and how.  
    \item \textbf{\RQ{2}:} \rqTwo
    We want to understand if developers would find beneficial changes aimed at fixing Dockerfile smells (\eg generated by an automated tool). 
\end{itemize}


\section{Study Context}

%

The context of our study is represented by samples of the dataset introduced by Eng \etal \cite{eng2021revisiting}. The dataset consists of about 9.4 million Dockerfiles, from a period of time spanning from 2013 to 2020. To the best of our knowledge, the dataset is the largest and the most recent one from those available in literature \cite{cito2017empirical,henkel2020learning,ksontini2021refactorings}. Moreover, it contains the change history (\ie commits) of each Dockerfile. This characteristic allows us to evaluate the survivability of code smells (\RQ{1}). The authors constructed that dataset thorough mining software repositories from the S version of the WoC (World of Code) dataset \cite{ma2019world}. From a total of 2 billions of commits and 135 million of distinct repositories, the authors extracted a total of about 9.4 million of Dockerfiles with a total of 11.5 million of unique commits. The final number of repositories is about 1.9 million. The dataset also contains the output of the \textit{hadolint} tool for each Dockerfile, that can be extracted from the replication package provided by Eng \etal \cite{eng2021replicationpkg} from their study.


\section{Execution Plan}

\begin{figure*}
    \centering
	\includegraphics[width=\linewidth]{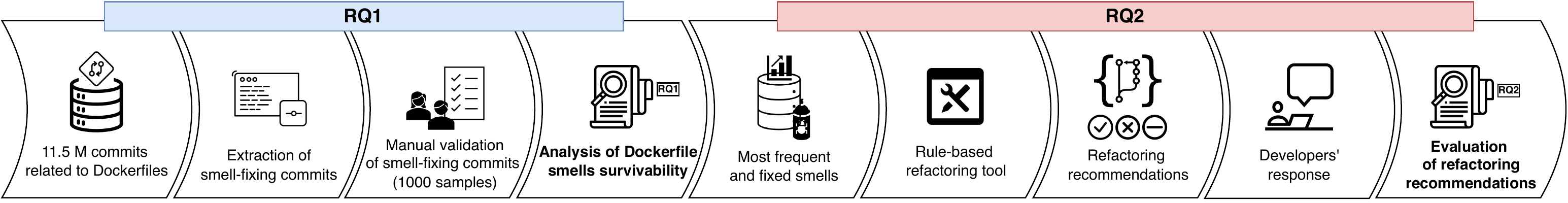}
	\caption{Overall workflow of the experimentation procedure.}
	\label{fig:workflow}
\end{figure*}

\begin{figure}[t]
    \centering
	\includegraphics[width=\linewidth]{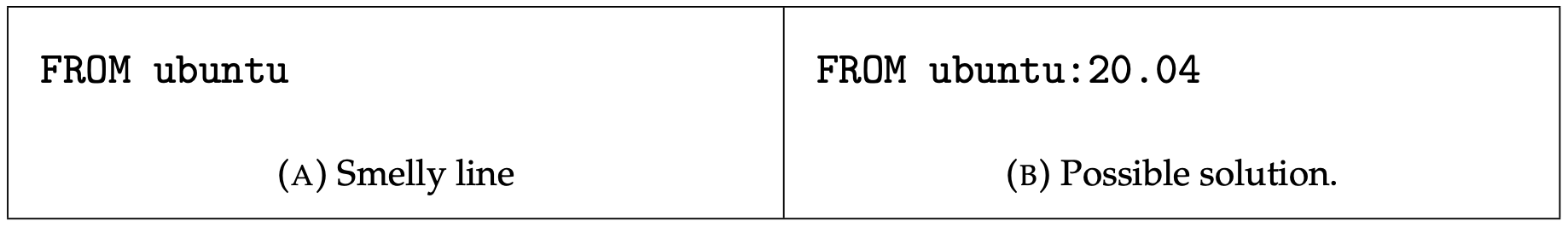}
	\caption{Example of rule DL3006.}
	\label{fig:dl3006}
\end{figure}


\begin{figure}[t]
    \centering
	\includegraphics[width=\linewidth]{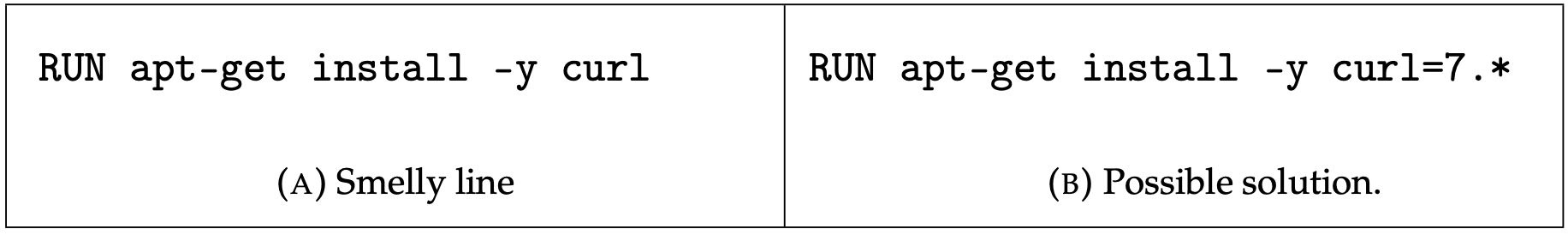}
	\caption{Example of rule DL3008.}
	\label{fig:dl3008}
\end{figure}

\begin{figure}[t]
    \centering
	\includegraphics[width=\linewidth]{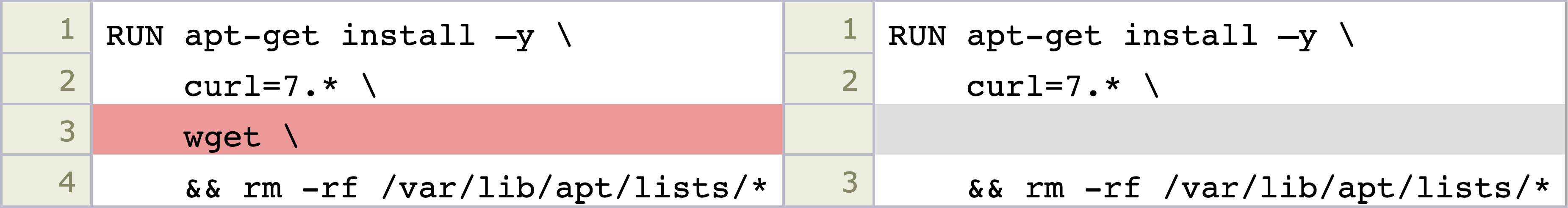}
	\caption{Example of a candidate smell-fixing commit that does not actually fix the smell.}
	\label{fig:invalidFix}
\end{figure} 

In this section, we describe the experimentation procedure that we will use to answer our RQs. \figref{fig:workflow} describes the overall workflow of the study.

\subsection{\RQ{1}: \rqOne}

To answer \RQ{1}, we will perform an empirical analysis on Dockerfile smell survivability. For each Dockerfile $d$, associated with the respective repository from GitHub, we will consider its snapshots over time, $d_1, \dots, d_n$, associated with the respective commit IDs in which they were introduced (\ie $c(d_1), \dots, c(d_n)$). We will also consider the Dockerfile smells detected with \textit{hadolint}, indicated as $\eta(d_1), \dots, \eta(d_n)$. For each snapshot $d_i$ (with $i > 1$) of each Dockerfile $d$, we will compute the disappeared smells as $\delta(d_i) = \eta(d_i) - \eta(d_{i-1})$. All the snapshots for which $\delta(d_i)$ is not an empty set are \textit{candidate} changes that aim at fixing the smells. We define a set of all such snapshot as $\mathit{PF} = \{d_i:|\delta(d_i)| > 0\}$.

As a next step, we want to ensure that the changes that led to the snapshots in $\mathit{PF}$ are actual fixes for the Dockerfile smell and if developers were aware of the smell when they made the change. 
To do this, we will manually inspect a sample of 1,000 of such candidate changes, which is statistically representative, leading to a margin of error of 3.1\% (95\% confidence interval), assuming an infinite population. 
We will look at the code diff to understand \textit{how} the change was made (\ie if fixed the smell or if the smell disappeared incidentally). Also, for actual fixes, we will consider the commit message, the possible issues referenced in it, and the pull requests to which they possibly belong to understand the purpose of the change (\ie if the fix was informed or not).
We will consider as a fix a change in which developers (i) modified one or more Dockerfile lines that contained one or more smells in the previous snapshot, and (ii) kept the functionality expressed in those lines. If, for example, the commit removes the instruction line where the smell is present. We will not label this as an actual smell-fixing commit because the smelly line is just removed and not fixed (\ie the functionality changed). Let us consider the example in \figref{fig:invalidFix}): The package \codeword{wget} lacks of version pinning (left). An actual fix would consist of the addition of a version to the package. Instead, in the commit, the package gets simply removed (\eg because it is not necessary). Therefore, we would not consider such a change as a fixing change.
We will mark a fix as \textit{informed} if the commit message, the possibly related pull request, or the issue possibly fixed with the commit explicitly report that the aim of the modification was to remove the smell actually removed.

At least two of the authors will independently evaluate each instance. The evaluators will discuss conflicts for both the aspects evaluated. In case of conflicts, the two evaluators will discuss aiming at reaching consensus. At the end, we will summarize the total number of fix commits and the percentage of actual fix commits. Moreover, for each rule violation, we will report the trend of smell occurrences and fixes over time, along with a summary table that describes the most fixed smells. We will also qualitatively discuss particular cases of fixing commits.

\begin{table*}[t]
	\centering
	\caption{The most frequent Dockerfile smells identified in literature \cite{eng2021revisiting}, along with the respective fixing rules we identified.}
	\resizebox{\linewidth}{!}{
		\begin{tabular}{l p{7cm} p{7cm}}
		\toprule
		\textbf{Smell name} & \textbf{Description} & \textbf{How to fix} \\
		\midrule
        DL3003  & Use \codeword{WORKDIR} to switch to a directory                       & Replace \codeword{cd} command with \codeword{WORKDIR} \\                   
        DL3006  & Base image version pinning                                            & Pin the version tag close to the Dockerfile commit date \\  
        DL3008  & apt-get version pinning                                               & Pin the software version close to the Dockerfile commit date \\                         
        DL3009  & Delete the apt-get lists after installing something                   & Add in the corresponding instruction block the lines to clean \codeword{apt} cache \\                                                     
        DL3015  & Avoid additional packages by specifying --no-install-recommends       & Add the option \codeword{--no-install-recommends} to the corresponding instruction block\\                                                                 
        DL3020  & Use \codeword{COPY} instead of \codeword{ADD} for files and folders   & Replace \codeword{ADD} instruction with \codeword{COPY} when copying files and folders \\                                                                    
        DL4000  & \codeword{MAINTAINER} is deprecated                                   & Replace maintainer with the equivalent \codeword{LABEL} instruction \\                                    
        DL4006  & Not using -o pipefail before \codeword{RUN}                           & Add the \codeword{SHELL} pipefail instruction before \codeword{RUN} that uses pipe \\                                             
		\bottomrule                   
		\end{tabular}
	}
	\label{tab:smells}

\end{table*}

\subsection{\RQ{2}: \rqTwo}

To answer \RQ{2}, we will first implement a tool for automatically fixing the most frequently occurring Dockerfile smells, based on a set of rules we defined. Then, we will use such a tool to fix smells in existing Dockerfiles from open-source projects and submit the changes to the developers through pull requests, to understand if they are keen to accept them.  
We describe such steps below.

\subsubsection{Fixing rules for Dockerfile Smells}

As a preliminary step, we identified a set of Dockerfile smells that we wanted to fix, considering the list of the most occurring Dockerfile smells, ordered by prevalence, according to the most recent paper on this topic \cite{eng2021revisiting}. However, we excluded and added some rule violations. Specifically, we excluded the rule violations DL3013 (\textit{Pin versions in pip}) and DL3018 (\textit{Pin versions in apk add}) because they are a less occurring variants (\ie ~4\% and ~5\%, respectively) of the more prevalent smell DL3008 (15\%), concerning different package managers. We report in \tabref{tab:smells} the full list of smells we will target in our study, along with the rule we will use to automatically produce a fix.
It is clear that most of the smells are trivial to fix. For example, to fix the violation DL3020, it is just necessary to replace the instruction \codeword{ADD} with \codeword{COPY} for files and folders. In the case of the \textit{version pinning}-related smells (\ie DL3006 and DL3008), instead, a more sophisticated fixing procedure is required. We refer to \textit{version pinning}-related smells as to the smells related to missing versioning of dependencies and packages. Such smells can have an impact on the reproducibility of the build since different versions might be used if the build occurs at different times, leading to different execution environments for the application. For example, when the version tag is missing from the \codeword{FROM} instruction of a Dockerfile (\ie DL3006), the most recent image having the latest tag is automatically selected. 
To fix such smells, we use a two-step approach: (i) we identify of the correct versions to pin for each artifact (\eg each package), and (ii) we insert the selected versions to the corresponding instruction lines in the Dockerfile. We describe below in more details the procedure we defined for each smell.

\textbf{Image version tag (DL3006).} This rule violation identifies a Dockerfile where the base image used in the \codeword{FROM} instruction is not pinned with an explicit tag. In this case, we use a fixing strategy that is inspired by the approach of Kitajima \etal \cite{kitajima2020latest}. 
Specifically, to determine the correct image tag, we use the image name together with the image \textit{digest}. 
Docker images are labeled with one or more \textit{tags}, mainly assigned by developers, that identify a specific version of the image when \textit{pulled} from DockerHub.
On the other hand, the \textit{digest} is a hash value that uniquely identifies a Docker image having a specific composition of dependencies and configurations, automatically created at build time.
The \textit{digest} of existing images can be obtained via the DockerHub APIs\footnote{\url{https://docs.docker.com/docker-hub/api/latest/}}. 
Thus, the only way to uniquely identify an image is using the \textit{digest}.
To fix the smell, we obtain (i) the \textit{digest} of the input Docker image through build, (ii) we find the corresponding image and its tags using the DockerHub APIs, and (iii) we pick the most recent tag assigned, if there are more than one, that not is the \textit{``latest"} tag. An example of smell fixed through this rule is reported in \figref{fig:dl3006}.

\textbf{Pin versions in package manager (DL3008).} The version pinning smell also affects package managers for software dependencies and packages (\eg \texttt{apt}, \texttt{apk}, \texttt{pip}). In that case, differently from the base image, the package version must be searched in the source repository of the installed packages. The smell regards the \textit{apt} package manager, \ie it might affect only the Debian-based Docker images. For the fix, we consider only the Ubuntu-based images since (i) we needed to select a specific distribution to handle versions (more on this later), and (ii) Ubuntu the most widespread derivative of Debian in Docker images \cite{eng2021revisiting}. 
The strategy we will use to solve DL3008 works as follows: First, a parser finds the instruction lines where there is the \texttt{apt} command, and it collects all the packages that need to be pinned. Next, for each package, a version number is selected considering the OS \textit{distribution} (\eg Ubuntu, Xubuntu, \etc), the series (\eg 20.04 \textit{Focal Fossa} or 14.04 \textit{Trusty Tahr}), and the \textit{last modification date} of the Dockerfile. The series of the OS is particularly important, because they may offer different versions for a same package. For instance, if we consider the \texttt{curl} package, we can have the version \texttt{7.68.0-1ubuntu2.5} for the \textit{Focal Fossa} series of Ubuntu, while for the series \textit{Trusty Tahr} it equals to \texttt{7.35.0-1ubuntu2.20}. So, if we try to use the first in a Dockerfile using the Trusty Tahr series, the build most probably will fail. In addition, the software package version having the closest date prior to the one that corresponds to the Dockerfile last modification is selected. The final step consists in testing the chosen package version. Generally, a package version adopts semantic versioning, characterized by a sequence of numbers in the format \texttt{$\langle MAJOR \rangle$.$\langle MINOR \rangle$.$\langle PATCH \rangle$}. However, the specific versions of the packages might disappear in time from the Ubuntu central repository, thus leading to errors while installing them. Given that the \texttt{PATCH} release does not drastically change the functionalities of the package and that old patches frequently disappear, we replace it with the symbol '*', indicating ``any version,'' where the latest will be automatically selected. After that, a simulation of the \texttt{apt-get install} command with the pinned version will be run to verify that the selected package version is available. If it is, the package can be pinned with that version; otherwise, also the \texttt{MINOR} part of the version is replaced with the ’*’ symbol. If the package can still not be retrieved, we do not pin the package, \ie we do not fix the smell. Pinning a different \texttt{MAJOR} version, instead, could introduce compatibility issues. 
It is worth saying that we apply our fixing heuristic only to packages having missing version pinning. This means that we do not update packages pinned with another version (\eg older than the reference date used to fix the smell).
Moreover, in some cases, developers may not want that pinned package version, but rather a different one. For example, they want a newer version of that package (\eg the latest).
We will evaluate that particular case in the context of \RQ{2}.
An example of fix generated through this strategy is reported in \figref{fig:dl3008}.

\subsubsection{Evaluation of Automated Fixes}

\begin{figure}[ht]
	\centering
	\begin{quote}
		\begin{summarybox}
			Hi!
	
			The Dockerfile placed at \placeholder{dockerfile\_path} contained a best practice violation, detected by the linting tool \textit{hadolint}, and identified as \placeholder{violation\_id}.
	
			The \placeholder{violation\_id} occurs when \placeholder{violation\_description} \\
				
			In this pull request, we propose a fix for the detected smell, automatically generated by a tool. To fix this smell, specifically, we \placeholder{fixing\_rule\_explanation}.
			This change is only aimed at fixing the specific smell. In case of rejection, please briefly indicate the reason (e.g., if you believe that the fix is not valid or useful and why, along with suggestions for possible improvement). \\
			
			Thanks in advance.
		\end{summarybox}
	\end{quote}
	\caption{Example of pull request message. The placeholders will be replaced with the corresponding value.}
	\label{fig:pull_request}
\end{figure}

We will propose the fixes generated by the tool we defined in the previous step to developers, so that we can evaluate if they are helpful. To achieve this, we will select a sample of fixes for each smell extracted from our dataset, in proportion of the total number of fixes. Moreover, we select at most one smell for each repository, to avoid flooding the developers with multiple pull requests. Also, to avoid toy projects, we select only Dockerfiles from repositories having at least 10 stars. We will perform a random stratified sampling, where we have the smell type as \textit{strata}. We will select a total of 384 instances, as it is sufficient to obtain a representative sample (5\% margin of error with 95\% confidence level, considering an unknown population size). Considering the smell occurrences reported by Eng \etal \cite{eng2021revisiting}, the less occurring is DL3006 with a percentage of 3.84\%. Considering that the total population is about 9.4M Dockerfiles, potentially there are approximately 352,000 instances having the DL3006 smell. Thus, we believe that we can easily obtain a sufficient number of instances for each \textit{strata} (\ie smell type) to perform the sampling. 

Next, for each fix in the selected sample, we will create a GitHub pull request where we propose to the developers the fix for the smell. We will use a GitHub account created specifically for this evaluation. We will select the Dockerfiles from repositories that have merged at least one pull request and have commit activity in the last three months. Also, the smell must still be present in the latest version of the Dockerfile. Finally, we discard all the Dockerfiles for which the build fails since we do not aim at fixing such a problem.

The pull request messages will have a body similar to the one described in \figref{fig:pull_request}.

We will adopt a methodology similar to the one used by Vassallo \etal \cite{vassallo2020configuration}. We will monitor the status of each pull request for 3 months, to allow developers to evaluate it and to give a response. We will interact with the developers if they ask questions or request additional information, but we will not make modifications to the source code of the proposed fix, unless they are strictly related to the smell (\eg the smell was not perfectly correct). We will explicitly mark those cases and report them. At the end of the monitoring period, each pull request can be in one of the following states: 

\begin{itemize}
    \item \textit{Ignored}: the pull request does not receive a response;
    \item \textit{Rejected/Closed}: the pull request has been closed or it is explicitly rejected;
    \item \textit{Pending}: The pull request has been discussed but it is still open;
    \item \textit{Accepted}: the pull request is accepted to be merged, but it is not merged yet;
    \item \textit{Fixed}: the proposed fix is in the main branch.
\end{itemize}

For each type of fixed smell, we will report the number and percentage of the fix recommendations accepted and rejected, along with the rationale in case of rejection and the response time. Also, we will conduct a qualitative analysis on the interactions of the developers with our pull requests. In particular, we will analyze those where the pull request is rejected or pending to understand why the fix was not accepted. For example, the fix is not accepted because it needs some modifications or adoption to a particular usage context, or else the developers simply are not interested in performing that modification to their Dockerfile. Moreover, we will evaluate the additional questions and information that the developer submit on both accepted and rejected pull requests. Two of the authors will use a card-sorting-inspired approach \cite{spencer2009card} on the obtained responses, where they will perform a first round of independent tagging, and then a second round of cross-validation to discuss conflicts. We will discuss and describe the resulting annotation and provide some lessons learned.


\section{Limitations, Challenges and Mitigations}

In this section we summarize the main limitations of our work, indicating the possible mitigation strategies to apply.

\textbf{Bias on Selected Smells.} There can be a bias in the selected smells for our fix recommendations. We selected the most occurring smell as described in the analysis of Eng \etal \cite{eng2021revisiting}. Our assumption is that an automated approach would have the biggest impact on the smells that occur more frequently. Also, at least for some of them, the reason behind the fact that they do not get fixed might be that they are not trivial to fix (\ie an automated tool would be most useful).

\textbf{Wrong Fixing Procedure for Rule violations.} The fixing procedure for the some of the selected smells can be wrong, and some smells might not get fixed. We based the rules behind the fixing procedure on the Docker best practices and on the \textit{hadolint} documentation. Still, to minimize the risk of this, we will double-check the modifications before submitting the pull requests and manually exclude the ones that make the build of the Dockerfile fail. It is worth noting, indeed, that our aim is not to evaluate the tool, but rather to understand if developers are willing to accept fixes.

\textbf{Not Enough Developers Interactions.} Considering the evaluation procedure involved in \RQ{2}, the worst case is that all the fix recommendations are ignored, thus leading to inconclusive results. To mitigate this risk, we will only select as target projects those that have at least one pull request accepted and a commit activity in the last three months. Also, we will submit a large number of pull requests (at most 384) to increase the likelihood of receiving a response.

\textbf{Effort for Handling Pull Requests.} In addition to that point, a large number of pull requests requires a lot of effort in monitoring of the developers' responses. We will implement a tool to monitor the pull requests we submitted and partially automatize such a task.


\section{Conclusion}

In the last few years, containerization technologies have had a significant impact on the deployment workflow. Best practice violations (\ie Dockerfile smells) are widely diffused in Dockerfiles \cite{cito2017empirical,wu2020characterizing,lin2020large,eng2021revisiting}. However, the scientific literature lacks studies aimed at understanding how developers fix such smells. We presented the plan for filling this gap. Our results will help researchers interested in supporting tools for improving code quality of Docker artifacts. We will also acquire qualitative feedback from developers, which will allow us to understand what are the benefits and the limitation of an automated tool for fixing Dockerfile smells. We will publicly release the results of our research (\ie the collected data and our tool for fixing Dockerfile smells) to foster future research in this field.

\bibliography{main}
\bibliographystyle{IEEEtran}

\end{document}